\documentclass[twocolumn,showpacs,preprintnumbers]{revtex4}
\usepackage{amssymb}
\usepackage{amsfonts}
\usepackage{amsmath}
\usepackage{graphicx}
\usepackage{dcolumn}
\usepackage{bm}

\input{tcilatex}

\begin{document}

\title{Dissipative dynamics of a charged particle in the field of three
plane waves: chaos and control}
\author{Ricardo Chac\'{o}n}
\affiliation{Departamento de Electr\'{o}nica e Ingenier\'{\i}a Electromec\'{a}nica,
Escuela de Ingenier\'{\i}as Industriales, Universidad de Extremadura,
Apartado Postal 382, E-06071 Badajoz, Spain}
\date{\today}

\begin{abstract}
The chaotic dissipative dynamics of a charged particle in the field of three
plane waves is theoretically (Melnikov's method) and numerically (Lyapunov
exponents) investigated. In particular, the effectiveness of one of such
waves in controlling the chaotic dynamics induced by the remaining two waves
is theoretically predicted and numerically confirmed. Two mechanisms
underlying the chaos-suppression scenario are identified. One mechanism
requires chaos-inducing and chaos-suppressing waves to have both
commensurate wavelengths and commensurate relative (with respect to the
remaining third wave) phase velocities, while the other mechanism allows the
chaotic dynamics to be tamed when such quantitites are incommensurate. The
present findings may be directly applied to several important problems in
plasma physics, including that of the chaos-induced destruction of magnetic
surfaces in tokamaks.
\end{abstract}

\maketitle

% Force line breaks with \\




The interaction of charged particles with a wave packet is a basic and
challenging problem appearing in many fundamental fields such as
astrophysics, plasma physics, and condensed matter physics [1,2], to name
just a few. As is well known, the structure of the wave packet reflects the
particular physical situation that is being considered in each case. In this
regard, a classical, widely studied particular case is that of an infinite
set of electrostatic waves having the same amplitudes and wave numbers, zero
initial phases, and integer frequencies [3]. While the Hamiltonian approach
to the general problem is suitable in many physical contexts [4], the
consideration of dissipative forces seems appropriate so as to take into
account such diverse phenomena as electron-ion collisions, energy losses
through synchrotron radiation, and stochastic heating of particles in plasma
physics. In any case, stochastic (chaotic) dynamics already appears (can
appear) when the wave packet solely consists of two plane waves [5-8]. As is
well known, this non-regular behavior of the charged particles may yield
undesirable effects in a number of technological devices such as the
destruction of magnetic surfaces in tokamaks [9]. Thus, in the context of
plasma physics \textit{inter al}., it is natural to consider the problem of
regularization of the chaotic dissipative dynamics of a charged particle in
a wave packet by a \textit{small} amplitude uncorrelated wave, which is
added to the initial wave packet. This Letter studies the simplest model
equation to examine this problem: 
\begin{eqnarray}
\overset{..}{x}+\gamma \overset{.}{x} &=&-\frac{e}{m}\left[ E_{0}\sin \left(
k_{0}x-\omega _{0}t\right) +E_{c}\sin \left( k_{c}x-\omega _{c}t\right) %
\right]   \notag \\
&&-\frac{e}{m}E_{s}\sin \left( k_{s}x-\omega _{s}t-\Psi \right) ,  \TCItag{1}
\end{eqnarray}%
where the amplitudes $E_{0},E_{c},E_{s},$ wave numbers $k_{0},k_{c},k_{s}$,
and frequencies $\omega _{0},\omega _{c},\omega _{s}$ correspond to the
\textquotedblleft main\textquotedblright ,\ chaos-inducing, and
chaos-suppressing waves, respectively, $\Psi $ is an initial phase, $e$ and $%
m$ are the charge and mass of the particle, respectively, and where \textit{%
weak} dissipation $\left( \gamma \ll 1\right) $ and \textit{non-uniform}
amplitudes $\left( E_{c,s}/E_{0}<1\right) $ are assumed. In a reference
frame moving along with the main wave, Eq. (1) transforms into the equation%
\begin{eqnarray}
\frac{d^{2}\xi }{d\tau ^{2}}+\sin \xi  &=&-\delta -\eta \frac{d\xi }{d\tau }%
-\varepsilon _{c}\sin \left( K_{c}\xi -\Omega _{c}\tau \right)   \notag \\
&&-\varepsilon _{s}\sin \left( K_{s}\xi -\Omega _{s}\tau -\Psi \right) , 
\TCItag{2}
\end{eqnarray}%
where $\xi \equiv k_{0}x-\omega _{0}t$, $\Omega _{0}\equiv \left(
ek_{0}E_{0}/m\right) ^{1/2}$, $\tau \equiv \Omega _{0}t$, $\delta \equiv
\gamma \omega _{0}\Omega _{0}^{-2}$, $\eta \equiv \gamma \Omega _{0}^{-1}$, $%
\varepsilon _{c,s}\equiv E_{c,s}/E_{0}$, $K_{c,s}\equiv k_{c,s}/k_{0}$, and $%
\Omega _{c,s}\equiv \left( \omega _{c,s}-\omega _{0}k_{c,s}/k_{0}\right)
\Omega _{0}^{-1}$ are all dimensionless variables and parameters. Also, $%
\Omega _{s}/\Omega _{c}=K_{s}v_{0s}/\left( K_{c}v_{0c}\right) =\lambda
_{c}v_{0s}/(\lambda _{s}v_{0c})$, where $v_{0c}\left( v_{0s}\right) $ and $%
\lambda _{c}\left( \lambda _{s}\right) $ are the relative phase velocity
with respect to the main wave $\left( v_{0c,s}\equiv \omega
_{c,s}/k_{c,s}-\omega _{0}/k_{0}\right) $ and the wavelength of the
chaos-inducing (-suppressing) wave, respectively. The parameters $%
k_{0},\omega _{0},E_{0},$ and $\Omega _{0}$ are held constant throughout.
Physically, Eq. (2) can be used to describe a number of important problems
in plasma physics: the dissipative dynamics of a charged particle in the
field of three electrostatic plane waves, the bounce dissipative motion of a
charged particle trapped in a toroidal magnetic field which is perturbed by
two (one chaos-inducing and the other chaos-suppressing) superposed
electrostatic waves, and the suppressory effect of a (secondary) resonant
magnetic perturbation on the destruction of magnetic surfaces by a (primary)
resonant magnetic perturbation in a tokamak. Since Eq. (2) represents a
perturbed pendulum $\left( 0<\delta ,\eta ,\varepsilon _{c},\varepsilon
_{s}\ll 1\right) $, one can apply Melnikov's method (MM) [10,11,2] to obtain
analytical estimates of the ranges of parameters $\left( E_{s},k_{s},\omega
_{s},\Psi \right) $ for suppression of the chaos existing in the absence of
the chaos-suppressing wave. The application of MM to Eq. (2) gives [12] the
Melnikov function%
\begin{equation}
M^{\pm }\left( \tau _{0}\right) =-D^{\pm }+A^{\pm }\sin \left( \Omega
_{c}\tau _{0}\right) +B^{\pm }\sin \left( \Omega _{s}\tau _{0}+\Psi \right) ,
\tag{3}
\end{equation}%
with $D^{\pm }\equiv 8\eta \pm 2\pi \delta $, $A^{\pm }\equiv 4\varepsilon
_{c}L^{\pm }\left( K_{c},\Omega _{c}\right) $, $B^{\pm }\equiv 4\varepsilon
_{s}L^{\pm }\left( K_{s},\Omega _{s}\right) $, and $L^{\pm }\left( K,\Omega
\right) \equiv \pm \int_{0}^{\infty }\func{sech}\tau \cos \left[ 2K\arctan
\left( \sinh \tau \right) \mp \Omega \tau \right] d\tau $ (see Fig. 1),
where the positive (negative) sign refers to the top (bottom) homoclinic
orbit of the underlying conservative pendulum. It is straightforward to
obtain the following properties: (\textit{i}) $L^{+}\left( K,\pm \Omega
\right) =-L^{-}\left( K,\mp \Omega \right) $; (\textit{ii}) $L^{+}\left(
K,\Omega \right) $ vanishes completely in the region $\left( K,\Omega
<0\right) $ but in a very narrow neighborhood of the line $\Omega =0$; (%
\textit{iii}) There exists a particular relationship, $\Omega =\alpha K$
with $\alpha \simeq \left( 1+\sqrt{5}\right) /2$ (i.e., the \textit{golden
ratio}), which determines the direction of the main tongue-like region where 
$L^{+}\left( K,\Omega \right) $ presents its \textit{largest} values (i.e.,
for small $K$ values); (\textit{iv}) $L^{\pm }\left( K=0,\Omega \right)
=\left( \pi /2\right) \func{sech}\left( \pi \Omega /2\right) $, $L^{\pm
}\left( K=n,\Omega =0\right) =0,n=1,2,...$ ; and (\textit{v}) $L^{\pm
}\left( K,\Omega \right) \in \left[ -\pi /2,\pi /2\right] $, $\forall \left(
K,\Omega \right) $. It is well known that the simple zeros of the Melnikov
function imply transversal intersection of stable and unstable manifolds,
giving rise to Smale horseshoes and hence hyperbolic invariant sets [11].
The mechanism for taming chaos by an added weak periodic perturbation is
then the frustration of a homoclinic (or heteroclinic) bifurcation [13].
Consider first the case with no chaos-suppressing wave $\left(
E_{s}=0\right) $. With fixed $\delta ,\eta ,\varepsilon _{c}$, one sees that
for \textit{sufficiently }small values of $k_{c}$ the best chance for the
occurrence of a homoclinic bifurcation, i.e., 
\begin{equation}
\frac{\left\vert 2\eta \pm \pi \delta /2\right\vert }{\varepsilon _{c}}%
<\left\vert L^{\pm }\left( K_{c},\Omega _{c}\right) \right\vert ,  \tag{4}
\end{equation}%
takes place when $\left\vert v_{0c}\right\vert $ is close to $\left\vert
v_{0c,\max }\right\vert $, where $v_{0c,\max }=\pm \alpha \Omega _{0}/k_{0}$
is the \textit{most chaotic} relative phase velocity (cf. properties (i) and
(iii)). Figure 2 shows an illustrative example of this prediction, where the
MM-based approximation (top panel) is compared with Lyapunov exponent (LE)
calculations of Eq. (2) (bottom panel). One sees that the largest values of
the maximal LE (black points) lie inside a narrow tongue-like region which
is near the theoretical estimate $\Omega =\alpha K$ (yellow line in Fig. 2).
Note that the maximal (positive) LE increases as $K_{c}$ is increased, which
is coherent with the corresponding growth of the chaotic threshold function $%
\left\vert L^{+}\left( K,\Omega \right) \right\vert $. With fixed $\delta
,\eta ,\varepsilon _{c}$, chaotic motion is possible when $\Omega _{c}=0$
(i.e., when the relative phase velocity vanishes) and $k_{c}\neq nk_{0},n$
positive integer (cf. property (\textit{iv}), although this possibility
decreases as $\lambda _{c}\rightarrow 0$. Also, as expected, property (i)
means that the chaotic threshold depends on the absolute value of the
relative phase velocity, but not on its \textit{sign} (cf. Eq. (4)). Let us
suppose in the following that, in the absence of any chaos-suppressing wave $%
\left( E_{s}=0\right) $, the associated Melnikov function $M_{0}^{\pm }(\tau
_{0})=-D^{\pm }+A^{\pm }\sin \left( \Omega _{c}\tau _{0}\right) $ changes
sign at some $\tau _{0}$, so that the charged particle exhibits (at least
transient) chaotic behavior. To study the taming effect of the
chaos-suppressing wave in the most favorable situation, the case of a
subharmonic ($\Omega _{s}=p\Omega _{c},p$ a positive integer) resonance is
analyzed below. For this case, one has%
\begin{equation}
\frac{v_{0s}/v_{0c}}{\lambda _{s}/\lambda _{c}}=p,  \tag{5}
\end{equation}%
which permits one to identify \textit{two different physical situations (or
mechanisms)} for the added chaos-suppressing wave to tame chaotic charged
particles: I. Chaos-inducing and chaos-suppressing waves having both \textit{%
commensurate} wavelengths ($\lambda _{s}/\lambda _{c}=m/n,m,n$ positive
integers) and \textit{commensurate} relative phase velocities ($%
v_{0s}/v_{0c}=m^{\prime }/n^{\prime },m^{\prime },n^{\prime }$ positive
integers) such that $p=m^{\prime }n/(n^{\prime }m)$. II. Chaos-inducing and
chaos-suppressing waves having both \textit{incommensurate} wavelengths and 
\textit{incommensurate} relative phase velocities such that the quotient $%
\left( v_{0s}/v_{0c}\right) /\left( \lambda _{s}/\lambda _{c}\right) $
satisfies Eq. (5).

Let $\Omega _{s}=p\Omega _{c}$, $p$ a positive integer, such that the
relationships $\Psi =\Psi _{opt}\equiv \pi $[$4m+3-p\left( 4n+1\right) $]$/2$%
, $\pi $[$4m+5-p\left( 4n+1\right) $]$/2$, $\pi $[$4m+3-p\left( 4n-1\right) $%
]$/2$, $\pi $[$4m+5-p\left( 4n-1\right) $]$/2$ are independently satisfied
for some positive integers $m$ and $n$ for the parameter regions $\left(
K,\Omega \right) $ where [$L^{\pm }\left( K_{c},\Omega _{c}\right) >0,L^{\pm
}\left( K_{s},\Omega _{s}\right) >0$], [$L^{\pm }\left( K_{c},\Omega
_{c}\right) >0,L^{\pm }\left( K_{s},\Omega _{s}\right) <0$], [$L^{\pm
}\left( K_{c},\Omega _{c}\right) <0,L^{\pm }\left( K_{s},\Omega _{s}\right)
>0$], and [$L^{\pm }\left( K_{c},\Omega _{c}\right) <0,L^{\pm }\left(
K_{s},\Omega _{s}\right) <0$], respectively. Then the frustration of a
homoclinic bifurcation occurs (i.e., $M^{\pm }\left( \tau _{0}\right) $
always has the same sign) if and only if the conditions%
\begin{eqnarray}
\varepsilon _{s,\min }^{\pm } &<&\varepsilon _{s}\leqslant \varepsilon
_{s,\max }^{\pm },  \notag \\
\varepsilon _{s,\min }^{\pm } &\equiv &\left( 1-\left\vert D^{\pm }/A^{\pm
}\right\vert \right) R^{\pm },  \notag \\
\varepsilon _{s,\max }^{\pm } &\equiv &R^{\pm }/p^{2},  \notag \\
R^{\pm } &\equiv &\varepsilon _{c}\left\vert L^{\pm }\left( K_{c},\Omega
_{c}\right) /L^{\pm }\left( K_{s},\Omega _{s}\right) \right\vert  \TCItag{6}
\end{eqnarray}%
are fulfilled for each region $\left( K,\Omega \right) $, respectively. (The
proof will be given elsewhere [12].)

Now one can make the following remarks. First, for the above $\left(
K,\Omega \right) $ regions, this result requires having ($\Psi =\pi ,\pi
/2,0,3\pi /2$), ($\Psi =0,3\pi /2,\pi ,\pi /2$), ($\Psi =0,\pi /2,\pi ,3\pi
/2$), ($\Psi =\pi ,3\pi /2,0,\pi /2$), respectively, for $%
p=4m-3,4m-2,4m-1,4m $ ($m=1,2,...$), respectively in each case. Second, the
effectiveness of the chaos-suppressing wave decreases as the resonance order 
$p$ is increased (cf. Eq. (6)). This is relevant in experimental
realizations of the control where the \textit{main} resonance case is then
expected to be the most favorable to reliably tame the chaotic dynamics.
Third, for the main resonance case $\left( \Omega _{s}=\Omega _{c}\right) $,
one has a more accurate and complete estimate [14] of the regularization
boundary in the $\Psi -\varepsilon _{s}$ parameter plane:%
\begin{equation}
\varepsilon _{s}^{\pm }=\left\{ \mp \cos \Psi \pm \sqrt{\cos ^{2}\Psi -\left[
1-\left( D^{\pm }/A^{\pm }\right) ^{2}\right] }\right\} R^{\pm },  \tag{7}
\end{equation}%
for $B^{\pm }>0$ and $A^{\pm }$, respectively, and%
\begin{equation}
\varepsilon _{s}^{\pm }=\left\{ \pm \cos \Psi \pm \sqrt{\cos ^{2}\Psi -\left[
1-\left( D^{\pm }/A^{\pm }\right) ^{2}\right] }\right\} R^{\pm },  \tag{8}
\end{equation}%
for $B^{\pm }<0$ and $A^{\pm }$, respectively. The two signs before the
square root apply to each of the sign superscripts of $\varepsilon _{s}^{\pm
}$, which, in its turn, is independent of the sign of $D^{\pm }$. Also, the
area enclosed by the boundary functions is given by $4\left( \left\vert
D^{\pm }/A^{\pm }\right\vert \right) R^{\pm }$. The relevance of the above
theoretical results on strange chaotic attractor elimination is confirmed by
means of LE calculations of Eq. (1) [15]. Figure 3 shows the results
corresponding to an illustrative example of the mechanisms of type I (top)
and type II for the main resonance case $\left( p=1\right) $. In the absence
of the chaos-suppressing wave $\left( \varepsilon _{s}=0\right) $, Eq. (1)
presents a strange attractor with a maximal LE $\Lambda ^{+}\left(
\varepsilon _{s}=0\right) =0.073\pm 0.001$ (bits/s). The diagrams in this
figure were constructed by only plotting points on the grid when the
respective LE was larger than $0$ (cyan points) or than $\Lambda ^{+}\left(
\varepsilon _{s}=0\right) $ (magenta points), and with the black dashed-line
contour denoting the theoretical boundary function (cf. Eq. (7)) which is
symmetric with respect to the optimal suppressory value $\Psi _{opt}=\pi $.
Regarding the mechanism of type I, one typically finds that complete
regularization $\left( \Lambda ^{+}\left( \varepsilon _{s}>0\right)
\leqslant 0\right) $ mainly appears inside the island which \textit{%
symmetrically} contains the theoretically predicted area, while
regularization by a type II mechanism seems to be \textit{insensitive} to
the initial phase $\Psi $. In this latter case, the lowest value $%
\varepsilon _{s,\min }^{+}$ of the theoretical boundary function roughly
coincides with the regularization threshold value. The regularization
regions in the $\Psi -\varepsilon _{s}$ parameter plane under the two types
of mechanisms present different features. For type I, one typically finds
that inside the regularization area which contains the predicted area, the
two non-null LEs are \textit{identical and constant} $\left( \Lambda
^{+}=\Lambda ^{-}=-\gamma /2\right) $, as is shown in the instance of Fig. 4
(top panel). This \textit{symmetry} property of the contraction of the phase
space volume for the predicted regularization areas in the $\Psi
-\varepsilon _{s}$ parameter plane is an inherent feature of the type I
mechanism. This property does not hold for the type II mechanism, since in
this case the regularization region is not completely "clean" (there exist
isolated points corresponding to chaotic behavior) and furthermore the
distribution of the non-null (negative) LEs is not perfectly uniform, as can
be appreciated in the instance of Fig. 4 (bottom panel).

In conclusion, the onset of chaotic dissipative dynamics of a charged
particle in the field of two plane waves and the effectiveness of an added
plane wave in taming that chaos have been theoretically predicted and
numerically confirmed. Two suppressory mechanisms were identified: One
mechanism requires chaos-inducing and chaos-suppressing waves to have both
commensurate wavelengths and commensurate relative phase velocities, while
the other allows chaos to be tamed when these quantities are incommensurate.
A more detailed discussion of the regularization routes by the two
mechanisms for the main and higher resonances $\left( p>1\right) $ as well
as of the Hamiltonian limiting case will be given elsewhere [12]. It should
be stressed that the results discussed in this work can be directly applied
to the problem of the chaos-induced destruction of magnetic surfaces in
tokamaks besides other important problems in plasma physics.

\bigskip

\bigskip

\bigskip

\bigskip

\bigskip

\bigskip

\bigskip

\bigskip

\bigskip

\bigskip

\bigskip

\bigskip

\bigskip

\bigskip

\bigskip

\bigskip

\bigskip

\bigskip

\bigskip

\bigskip

\bigskip

\bigskip

\bigskip

\bigskip

\textbf{Figure Captions}

\bigskip

Figure 1 (color online). Function $L^{+}\left( K,\Omega \right) $ (see the
text).

\bigskip

Figure 2 (color online). Top panel: Contour plot of the threshold function $%
\left\vert L^{+}\left( K,\Omega \right) \right\vert $ with a gray scale from
white (1.0 contour) to black (0.0 contour). Bottom panel: Grid of $200\times
200$ points in the $K_{c}-\Omega _{c}$ parameter plane where cyan, magenta,
and black points indicate that the respective LE was larger than 0.0, 0.07,
and 0.14, respectively. System parameters: $\delta =\eta =0.1,\varepsilon
_{c}=0.7$.

\bigskip

Figure 3 (color online). Grids of $100\times 100$ points in the $\Psi
-\varepsilon _{s}$ parameter plane for the mechanisms of type I (top panel, $%
k_{s}=k_{c},v_{0s}=v_{0c}$) and type II (bottom panel, $k_{s}=\left( \sqrt{5}%
-1\right) k_{c}/2,v_{0s}=\left( \sqrt{5}+1\right) v_{0c}/2$. Black
dashed-line contour indicates the predicted boundary (cf. Eq. (7)). System
parameters: $k_{0}=\omega _{0}=\Omega _{0}=1,\Omega _{s}=\Omega
_{c},\varepsilon _{c}=0.7,k_{c}=1.22,\omega _{c}=2.26,\gamma =0.1$.

\bigskip

Figure 4. Top panel: Maximal LE $\Lambda ^{+}$ vs the initial phase $\Psi $
for six values of the suppressory amplitude $\varepsilon
_{s}=0.2,0.4,0.6,0.8,1.0,1.2$ corresponding to the type I mechanism. Bottom
panel: Maximal LE $\Lambda ^{+}$ vs suppressory amplitude $\varepsilon _{s}$
for six values of the initial phase $\Psi =\pi /3,2\pi /3,\pi ,4\pi /3,5\pi
/3$ corresponding to the type II mechanism. Dashed-line contour indicates
the predicted boundary (cf. Eq. (7)). System parameters as in the caption to
Fig. 3.

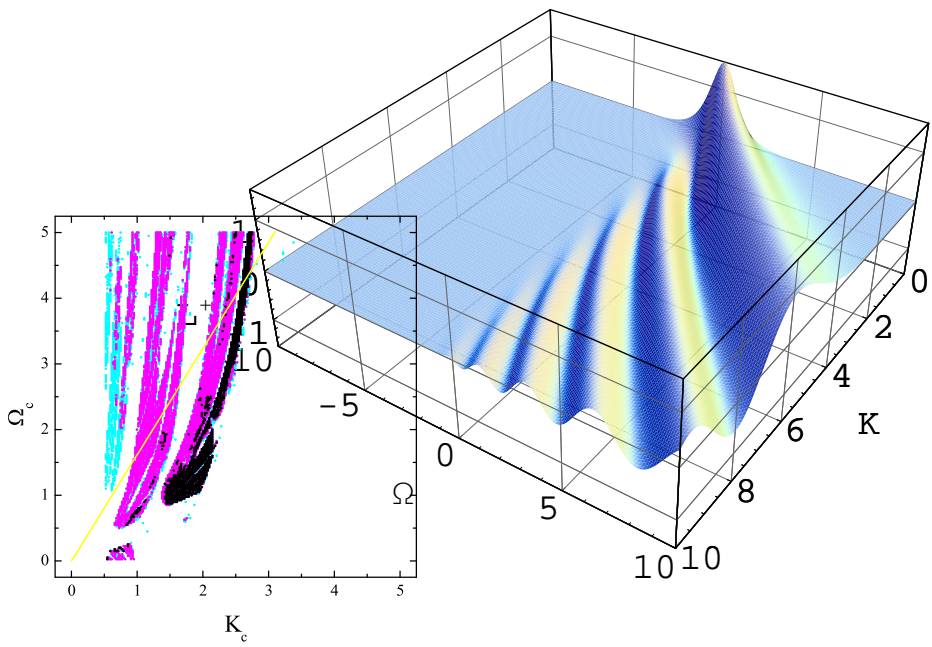
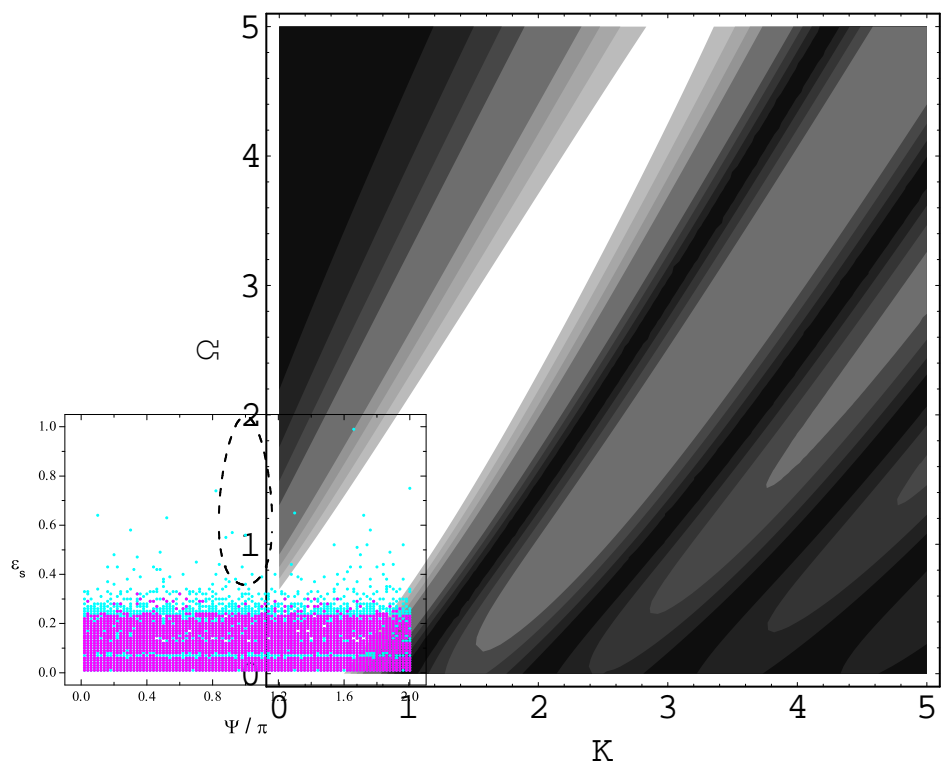
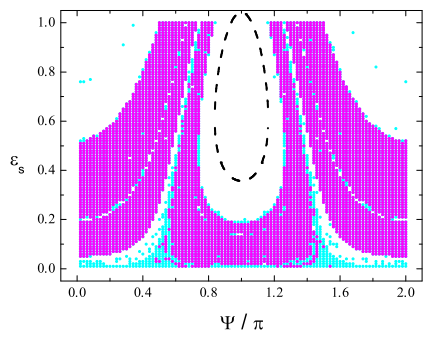
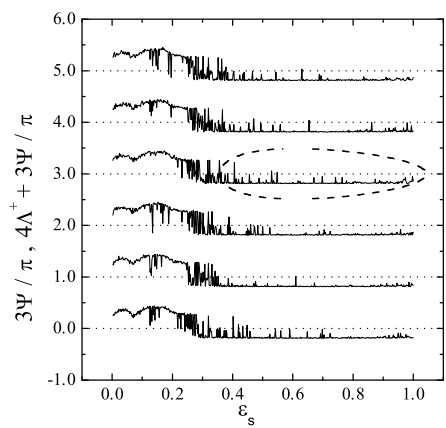
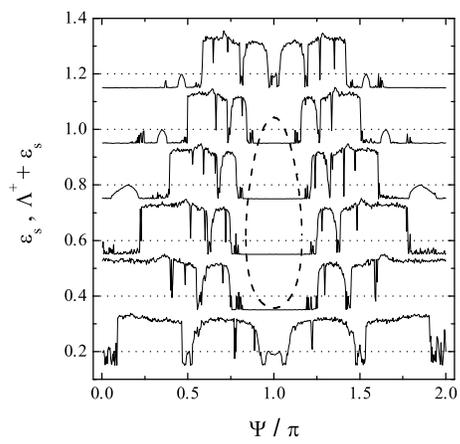
\end{document}